\documentclass[a4paper,fleqn,usenatbib]{mnras}
\usepackage[T1]{fontenc}
\usepackage{ae,aecompl}

\usepackage{graphicx}	
\usepackage{amsmath}	
\usepackage{amssymb}	

\def\apjs{{\rm ApJS}}                  
\def\apjl{{\rm ApJ Let}}               
\def\mnras{{\rm MNRAS}}                        

\def\Msol{\thinspace\hbox{$\hbox{M}_{\odot}$}}

\def\a4{\hsize 17.0cm \vsize 25.cm}
\newcommand{\der}[2]  { \frac{{\rm d}#1}{{\rm d}#2} }

\title[Gas expulsion vs gas retention in YSCs]{Gas expulsion 
        vs gas retention: what process dominates in young massive clusters?}

\author[S. Silich and G. Tenorio-Tagle]
{Sergiy Silich \thanks{E-mail: silich@inaoep.mx}
and Guillermo Tenorio-Tagle
\\
Instituto Nacional de Astrof\'\i sica \'Optica y Electr\'onica, AP 51, 
      72000 Puebla, M\'exico\\
}

\date{Accepted XXX. Received YYY; in original form ZZZ}

\pubyear{2016}

\begin{document}
\label{firstpage}
\pagerange{\pageref{firstpage}--\pageref{lastpage}}
\maketitle

\begin{abstract}
The ability of young stellar clusters to expel or retain the gas left over 
after a first episode of star formation is a central issue in all models 
aiming to explain multiple stellar populations and the peculiar light element 
abundance patterns in globular clusters. Recent attempts to detect the gas 
left over from star formation in present day clusters with masses similar to
those of globular clusters did not reveal a significant amount of gas 
in the majority of them, which strongly restricts the scenarios of multiple 
stellar population formation. Here the conditions required to retain the gas 
left over from star formation within the natal star forming cloud are revised.
It is shown that the usually accepted concept regarding the thermalization of 
the star cluster kinetic energy  due to nearby stellar winds and SNe ejecta 
collisions must be taken with care in the case of very compact and dense star 
forming clouds where three star formation regimes are possible if one 
considers different star formation efficiencies and mass concentrations. 
The three possible regimes are well separated in the half-mass radius and
in the natal gas central density vs pre-stellar cloud mass parameter space.  
The two gas free clusters in the Antennae galaxies and the gas rich cluster 
with a similar mass and age in the galaxy NGC 5253 appear in different zones 
in these diagrams. The critical lines obtained for clusters with a solar and 
a primordial gas metallicity are compared. 
\end{abstract}


\section{Introduction}
\label{sec1}

Globular clusters (GCs), considered for a long time to be chemically 
homogeneous stellar systems, formed instantaneously in young assembling 
galaxies, are now confirmed to have a much more complex structure. Deep 
photometric observations and detailed spectroscopy with Hubble Space Telescope
have revealed that all of them have distinct main sequence subpopulations and
large star-to-star abundance variations in light elements with well documented
anti-correlations between O and Na, Mg and Al 
\citep[see][and references therein]{Bedin2004,Marino2008,Carretta2009,
Piotto2012,Renzini2015}.
At the same time, the majority of GCs remain mono-metallic regarding the iron 
group elements. Only the most massive ones (e.g. $\omega$ Cen, M22, Terzan 5) 
exhibit multiple [Fe/H] populations which strongly restricts the ability of 
young proto-globular clusters to retain the iron enriched supernovae (SNe) 
ejecta \citep[see][and references therein]{Renzini2013}.
Despite many efforts to solve this problem, the discovery of multiple 
stellar populations in globular clusters remains one of the most challenging 
results of the last decade in the context of the origin and evolution of 
stellar populations in galaxies. 

There are three main candidates that potentially reach the right hydrogen 
burning conditions to explain the abundance anomalies in globular clusters.
These are: a) asymptotic giant branch (AGB) stars \citep[][]{DAntona2004,
DErcole2008}; b) fast rotating massive stars \citep[FRMS,][]{Prantzos2006,
Decressin2007,Bastian2013} and interacting massive binaries 
\citep[IMB,][]{DeMink2009,TenorioTagle2016}.

The applicability of the FRMS and IMB scenarios requires that young 
proto-globular clusters retain some of the pristine gas left over from the 
formation of a first stellar population for some time (20Myr-40Myr) 
whereas in the AGB scenario this gas is removed by stellar winds and SN 
explosions. The second stellar population in this latter case is formed after 
the end of SN explosions from the slow AGB winds retained in the potential 
well of the cluster mixed with some matter accreted late in the evolution
of the cluster. 

Recently \citet{Bastian2013,Bastian2014} attempted to find the matter left 
over from star formation in nearby young massive clusters, but did not 
detect any significant amount of gas (down to 1-2 per cent limits of the 
star cluster mass) in their target clusters. This led them 
to claim that even high-mass clusters with masses $10^6$\Msol \, or more, 
expel their natal gas within a few Myr after their formation, which 
strongly restricts scenarios for multiple stellar population formation.
However observations of nearby galaxies provided in infrared, millimeter and 
radio wavelengths revealed several very luminous and compact regions which are
believed to be young stellar clusters still embedded into their natal gaseous
clouds \citep[][]{Turner2000,Turner2003,Beck2015}.

1D and 3D numerical simulations provided by \citet{DErcole2008} and 
\citet{Calura2015} show that stellar winds
effectively remove gas from star forming regions and completely clean them up 
of the pristine matter no later than 15Myr after their formation.  
However calculations provided by \citet{Krause2012,Krause2016} with a 1D 
thin layer approximation led to the conclusion that in most cases the 
expanding star cluster wind-blown shells are destroyed via Rayleigh-Taylor 
(RT) instabilities before reaching the escape velocity which prevents gas 
expulsion from the star forming region.

Another approach has been recently discussed by \citet{TenorioTagle2015,
TenorioTagle2016}
who considered a model in which the negative feedback provided by stellar 
winds is suppressed because all massive stars compose close binary systems. 
The major implications of this are that the large collection of interacting 
massive binaries are likely to hold the remaining cloud against gravitational 
collapse while contaminating the gas left over from star formation. This would
happen without disrupting the centrally concentrated density distribution 
even during the early evolution of globular clusters. Under such conditions, 
blast waves from sequential supernova explosions are likely to undergo 
blowout expelling the SN products into the ambient interstellar medium (ISM) 
keeping most clusters mono-metallic regarding the iron group elements.
     
Here we show that bubbles formed by stellar winds and single SN explosions in 
the central zones of compact and massive young stellar clusters may stall 
before merging with their neighbors. This naturally allows for the retention 
and enrichment of the gas left over from the formation of a 1G and likely 
would lead to the formation of a second generation of stars with a peculiar 
abundance pattern. The fate of the gas left over from star formation depends 
then not only on the proto-stellar cloud mass and the star formation 
efficiency (SFE), but also on the gas concentration (or the central gas 
density) in the proto-stellar cloud. The major output from our study is 
critical lines which separate the gas expulsion and the gas retention regimes 
in the star forming cloud size and in the central gas density vs the star 
forming cloud mass parameter space and the dependence of these critical lines 
on the SFE.

Radiative pressure effects are presumably small in the dynamics of bubbles 
formed around single OB stars \citep{Krumholz2009}. They may be 
significant only for a short while in the dynamics of shells formed around 
star clusters \citep[e.g.][]{Silich2013,SergioMG2014,Gupta2016}. Such shells 
are driven by a high thermal pressure, which is boosted when individual 
stellar winds begin to collide and convert their kinetic energy into thermal 
energy of the injected matter. We leave the detailed analysis of radiation 
pressure effects on our model to a forthcoming communication.  

The paper is organized as follows. In section 2 we introduce the models 
adopted for the initial gas and stellar density distributions in the star
forming cloud and determine the rate of mechanical energy injected by massive
stars into the intra-cluster medium. In this section we also discuss the 
dynamics of bubbles driven by individual stellar winds and individual 
supernova explosions and 
derive the conditions for them to stall before merging with their neighbors. 
In section 3 we split the half-mass radius vs star forming cloud mass and the 
central gas density vs star forming cloud mass parameter space onto three 
regions. In the first case, neighboring wind-driven bubbles merge and 
upon the collisions their shells fragment leaving the swept-up matter as a
collection of dense clumps. Neighboring stellar winds begin then to interact
directly causing the large overpressure that evolves into a cluster wind.
The fragmented original gas will then be mass-loaded by the cluster wind and 
finally expelled out of the star forming volume. In the second case the 
wind-driven bubbles are unable to merge and are thus not able to cause the 
conditions for the generation of a cluster wind and the removal of the 
leftover gas. In this regime individual SNe may engulf neighboring wind 
sources. 
The energy input rate to the SNR then grows with time causing a continuous 
acceleration and eventually the SNR destruction. This immediately leads to 
the expelling of SN products out of the cluster. In the third region SN 
remnants stall before merging with nearby wind-driven bubbles. The natal gas 
is then contaminated with the iron group elements and retained in the central 
zones of the star forming cloud. Here we also discuss how the SFE and the 
stellar metallicity affect the critical lines which separate clusters 
evolving in different regimes. We also show that three clusters with similar 
masses and ages, two gas-less clusters in the Antennae galaxies and the gas 
rich cluster in the dwarf galaxy NGC 5253, are located in different regions of
our half-mass radius vs star forming cloud mass diagram. Finally, in section 4
we summarize our major results and conclusions.  

\section{Model setup}
\label{sec2}

\subsection{Stellar and gas density distributions}

In order to study the feedback that recently formed stars provide on their
parental gaseous cloud one has to select a model for the mass density 
distribution in the star-forming region. Following recent works 
\citep[e.g.][]{Krause2012,Calura2015} we adopt a Plummer density distribution: 
\begin{eqnarray}
      \label{eq.1a}
      & & \hspace{-1.1cm} 
\rho_g(r) = \frac{3 M_g}{4 \pi a^3} \left(1 + \frac{r^2}{a^2}\right)^{-5/2} ,
      \\[0.2cm] \label{eq.1b}
      & & \hspace{-1.1cm} 
n_{\star}(r) = \frac{3 N}{4 \pi a^3} \left(1 + \frac{r^2}{a^2}\right)^{-5/2} ,
\end{eqnarray}
where $\rho_g$ and $n_{\star}$ are the gas volume density and the
massive star number density in the star forming cloud, respectively, 
$M_g = (1 - \epsilon M_{tot})$ is the total amount of gas left over from the 
formation of the first stellar generation, $M_{tot}$ is the total mass of the 
star forming cloud, $\epsilon$ is the efficiency of star formation and $N$
is the total number of massive stars, the major feedback agent in any young
massive cluster. In the case of a Plummer density distribution the half-mass
radius is $R_{hm} = 1.3 a$, where $a$ is the characteristic length scale for 
the gaseous and the stellar density distributions. We assume a standard 
Kroupa initial mass function (IMF). The number of single massive stars $N$
with $M > 8$\Msol \, then scales with the star cluster mass as 
\citep[e.g.][]{Calura2015}:
\begin{equation}
      \label{eq2}
N = N_0 (M_{\star}/10^6\Msol)
\end{equation}
where $N_0 = 10^4$, $M_{\star} = \epsilon M_{tot}$. The mean separation
between nearby massive stars, $\Delta = 2 X$, is determined by the condition
that within a sphere of radius $X$ there is only one massive star:
$4 \pi X^3 n_{\star} / 3 = 1$. In the case of the Plummer density distribution
$X$ grows with distance to the star cluster center:
\begin{equation}
      \label{eq3}
X = a  \left(1 + \frac{r^2}{a^2}\right)^{5/6} / N^{1/3} . 
\end{equation}

\subsection{Input energy}

Another input parameter is $L_{\star}$ - the star cluster mechanical 
luminosity. $L_{\star}$ changes in stellar populations with different 
metallicities and in recent starbursts is expected to be larger than it was 
at the epoch when the first stellar clusters (globular clusters) formed. 
$L_{\star}$ scales with the star cluster mass as
\begin{equation}
      \label{eq4}
L_{\star} = L_0 (M_{\star}/10^6\Msol) ,
\end{equation}
where the normalization coefficient $L_0 \approx 3 \times 10^{40}$erg
s$^{-1}$ for young stellar clusters with solar metallicity, standard 
Kroupa IMF and Padova stellar evolutionary tracks \citep[][]{Leitherer1999}.
For stellar clusters with a low metallicity, analogues to proto-globular
clusters, we use the \citet{Calura2015} prescription for the stellar wind 
power which results in an order of magnitude smaller star cluster 
mechanical luminosity and thus reduces the value of the normalization 
coefficient $L_0$ in equation (\ref{eq4}) to $L_0 = 3 \times 10^{39}$~erg 
s$^{-1}$. The typical power of individual stellar winds then is $L_s = 
L_{\star}/N$. For supernovae explosions a standard value of $10^{51}$erg was 
adopted regardless of the star formation epoch.

Note that the number of massive stars $N$ decreases after the onset of SN 
explosions (it is assumed that the rate of SN explosions follows the 
Starburst99 model predictions). The mean separation between nearby massive 
stars $\Delta$ then increases with time. Clusters which were not able to form
a wind and expel the leftover gas after the first SN explosion should 
retain it when the separation between nearby energy sources becomes larger. 
Therefore hereafter we do not consider the star cluster mechanical luminosity
time evolution.

The results obtained below should not be restricted to the Plummer model.
For example, in the case of mass segregation  \citep[e.g.][]{Dib2008} the
assumption that the gas and massive stars are equally distributed does not
hold.  The mean separation between massive stars in the central zones of the 
cluster then would be smaller than that predicted by equation (\ref{eq3}).
This will reduce the critical radii obtained in the next sections and enhance
the gas critical densities, but does not change our major conclutions.

\subsection{Wind-driven bubbles}
\label{sec2}

Each massive star composes a wind which sweeps up the surrounding medium
into a shell and forms a bubble filled with a hot shocked wind gas. If 
thermal conduction and mass evaporation from the wind-driven shell are 
inhibited by magnetic fields the radiative losses of energy from the shocked 
wind region are negligible \citep[][]{Silich2013}. Wind-driven bubbles then 
evolve in the snow-plow regime and the thermal pressure inside of the 
wind-driven bubble decreases as the bubble grows larger 
\citep[see][]{Bisnovatyi1995}:
\begin{equation}
      \label{eq4a}
P = \frac{7}{25} \left[\frac{375(\gamma-1) 
                 L_{\star}}{28(9\gamma-4)\pi N}\right]^{2/3}
                 \rho_g^{1/3} r^{-4/3} .
\end{equation}

The bubble stalls when this pressure drops to the pressure in the ambient 
intra-cluster medium. We assume that in the central zones of star-forming 
clouds the turbulent pressure dominates over the thermal one.  The 
stalling radius is then determined by the condition that $P = P_t$, where the 
turbulent pressure $P_t = \rho_g \sigma^2$ \citep[e.g.][]{Smith2006}.
The one-dimensional velocity dispersion $\sigma$ is
\begin{equation}
      \label{eq5}
\sigma = \left(\frac{G M_{tot}}{\beta R_{hm}}\right)^{1/2} ,
\end{equation}
where $G$ is the gravitational constant and $R_{hm}$ is the half-mass radius
of the star forming cloud. In the case of a spherical cluster with an 
isotropic velocity distribution $\beta = 7.5$ \citep[][]{Smith2001}.

One can calculate now where the bubble around a single massive star stalls:
\begin{eqnarray}
      \nonumber
      & & \hspace{-1.1cm} 
R_{stall,w} = \left(\frac{7 \beta}{25 G}\right)^{3/4} 
              \left[\frac{125(\gamma-1) L_{\star}}
               {7 (1.3)^3 (9\gamma-4)(1-\epsilon) N}\right]^{1/2} \times
      \\[0.2cm]     \label{eq6}
      & & \hspace{0.5cm}
\left(\frac{1 + r^2/a^2}{M_{tot}}\right)^{5/4} R^{9/4}_{hm} .
\end{eqnarray}
Stellar winds do not merge if the stalling radius is smaller than 1/2 the 
distance between nearby sources: $R_{stall,w} < X$. This condition
together with equations (\ref{eq3}) and (\ref{eq6}) allow one to determine
how compact the parental cloud should be in order to keep newly born 
stars buried within the pristine gas left over from star formation: 
\begin{eqnarray}
      \nonumber
      & & \hspace{-1.1cm} 
R_{hm} < \left(\frac{25 G}{7 \beta}\right)^{3/5}
         \left[\frac{9.1 (9\gamma-4)(1-\epsilon) N}               
               {125(\gamma-1) L_{\star}}\right]^{2/5} \times
      \\[0.2cm]     \label{eq7}
      & & \hspace{0.15cm}
               \frac{M_{tot}}{N^{4/15} (1 + r^2/a^2)^{1/3}} .
\end{eqnarray}
$R_{hm}$ in equation (\ref{eq7}) is the half-mass radius of the star forming 
cloud and $r$ is the size of the central zone where stellar winds do not 
merge. The condition that $r = 0$ determines then the critical half-mass 
radius and the critical central density in the star forming cloud:
\begin{eqnarray}
      \nonumber
      & & \hspace{-1.1cm} 
R_{hmw,crit}=\left(\frac{25 G}{7 \beta}\right)^{3/5}
               \left[\frac{9.1 (9\gamma-4)(1-\epsilon) N}
               {125(\gamma-1) L_{\star}}\right]^{2/5} \times
      \\[0.2cm]     \label{eq8a}
      & & \hspace{0.7cm}
               \frac{M_{tot}}{N^{4/15}} ,
      \\[0.2cm]     \label{eq8b}
      & & \hspace{-1.1cm}
\rho_{w,crit}=\frac{3 M_g}{4 \pi a^3_{crit}} = 
              \frac{3 (1 - \epsilon) M_{tot}}{4 \pi (R_{hmw,crit}/1.3)^3} .
\end{eqnarray}
If the half-mass radius of the star forming cloud is larger than
$R_{hmw,crit}$ and thus the central density is smaller than the critical one, 
stellar winds merge everywhere inside the star forming volume. Their 
merging results in the fragmentation of individual wind-driven shells which 
leads to a plethora of dense clumps around the sources and at the same time 
allows for the direct collision and thermalization of neighboring winds. 
The hot gas then rapidly streams away from the center comprising a cluster
wind which can be mass-loaded with the fragmented matter 
\citep[e.g.][]{Silich2010,Rogers2013}. 
Such a wind eventually cleans up the star forming volume from the natal gas and
prevents a 2G formation \citep{Calura2015}.
However, if the half-mass radius of the proto-cluster cloud is smaller than 
$R_{hmw,crit}$ and thus the central density is larger than $\rho_{w,crit}$, 
stellar winds do not merge in the central zone of the cluster. The size of 
this zone $R_{2G}$ depends on the actual half-mass radius of the proto-stellar
cloud and becomes larger as one considers clouds with radii smaller than 
$R_{hmw,crit}$.

Inside the central zone with radius $r < R_{2G}$ the 1G stars contaminate the 
pristine gas left over after their formation with different H-burning products
providing the conditions to form a polluted 2G. Note that wind-driven bubbles 
do not stall in the outskirts of the cloud where they instead accelerate into 
a sharp density gradient and then blow out into the surrounding ISM. This 
restricts the size of the polluted zone and results in a centrally 
concentrated second subpopulation.

\subsection{Supernovae-driven bubbles}

In young stellar clusters with a normal IMF stellar winds represent a dominant
negative feedback mechanism only for a short
while, unless the most massive stars do not explode as supernovae, but directly
form black holes \citep[see][and references therein]{Decressin2010,Krause2012}.
It is usually assumed that in massive star clusters the first supernova 
explodes at an age of about 3.5Myr and since then supernova explosions 
become the most energetic negative feedback events responsible for the gas 
expulsion from the recently formed clusters. In this section we discuss the 
impact that SNe provide on the gas which a cluster wind did not expel from the 
star forming region and the conditions required to form a second 
subpopulation enriched with the iron group elements.

If the gas density inside a star forming cloud exceeds the critical value 
determined by equation (\ref{eq8b}), supernova remnants (SNRs) may stall 
either before or after merging with nearby wind-blown bubbles. As we show 
in the next section, the critical densities calculated by means of equation 
(\ref{eq8b}) are large. In such a case the Sedov phase terminates very rapidly
\citep[][]{Shull1980,Wheeler1980} and supernova remnants evolve 
in the snow-plow regime. The thermal pressure inside the supernova remnant 
then drops as it grows \citep[][]{Pasko1986}:
\begin{equation}
      \label{eq10}
P = \frac{3(\gamma-1)}{4 \pi} E_0 R^{3(\gamma-1)}_0 r^{-3\gamma} ,
\end{equation}
where $R_0$ is the radius at which the transition to the snow-plow
regime has occurred and $E_0$ is the thermal energy of the remnant
at this moment. The remnant stops after this pressure drops to 
that in the turbulent ambient medium which leads to the SN-driven bubble 
stalling radius 
\begin{eqnarray}
      \nonumber
      & & \hspace{-1.1cm} 
R_{stall,SN} = \frac{1}{1.3^{1/\gamma}} \left[\frac{\beta (\gamma-1) 
               E_0 R^{3(\gamma-1)}_0}{(1-\epsilon) G M^2_{tot}}\right]^
               {1/3 \gamma} \times
      \\[0.2cm]     \label{eq11}
      & & \hspace{0.5cm}
\left(1 + \frac{r^2}{a^2}\right)^
               {5/6\gamma} R_{hm}^{4/3 \gamma} .
\end{eqnarray}

As the SN expands into the rarefied gas left by a previous stellar wind,
the initial radius $R_0$ could be approximated by the radius of the stalling
wind-driven bubble: $R_0 \approx R_{stall,w}$ \citep[][]{Wheeler1980}. 

Numerical simulations \citep[e.g.][]{Chevalier1974,Falle1981,TenorioTagle1990}
show that about 1/2 of the supernova remnant thermal energy is radiated away 
during a short transition phase from the adiabatic to the radiative regime. 
Therefore it was assumed that the initial thermal energy $E_0$ is
\citep[][]{Bisnovatyi1995}:
\begin{equation}
      \label{eq12a}
E_0 = \frac{1}{2} \frac{\gamma + 1}{3\gamma -1} E_{SN} ,
\end{equation}
where $E_{SN} = 10^{51}$erg is the supernova explosion energy. Note that in
the very dense ambient medium ($n > 10^5$cm$^{-3}$), strong radiative cooling
sets in before the Sedov phase starts and the initial energy $E_0$ could 
be even smaller than this value \citep[][]{Terlevich1992}. 

The SN remnant stalling radii are much larger than those of wind-driven 
bubbles expanding into an equally dense ambient medium. This allows one to 
obtain the half-mass radii required for SNR to stall before merging with 
nearby wind-blown bubbles from the condition that $R_{stall,SN} < 2 X$:
\begin{eqnarray}
      \nonumber
      & & \hspace{-1.1cm} 
R_{hm} < \left(\frac{2}{N^{1/3}}\right)^{\frac{12\gamma}{(15\gamma-11)}}
   \left(\frac{25 G}{7 \beta}\right)^{\frac{9(\gamma-1)}{(15\gamma-11)}}
   \times
      \nonumber      
          \\[0.2cm]      
          & & \hspace{-1.1cm} 
   \left[\frac{9.1 (9\gamma-4)(1 - \epsilon) N}
   {125(\gamma-1) L_{\star}}\right]^{\frac{6(\gamma-1)}{(15\gamma-11)}}
   \left[\frac{(1-\epsilon) G M_{tot}}
   {\beta (\gamma-1) E_0}\right]^{\frac{4}{(15\gamma - 11)}}
   \times
      \nonumber
          \\[0.2cm]      
          & & \hspace{-1.1cm} 
   \left(1 + \frac{r^2}{a^2}\right)^{-\frac{5(\gamma-1)}{(15\gamma-11)}} 
    M_{tot} .
      \label{eq12}
\end{eqnarray}
The condition that $r = 0$ in equation (\ref{eq12}) determines the critical 
half-mass radius and the critical central density in the star forming cloud:
\begin{eqnarray}
      \nonumber
      & & \hspace{-1.1cm} 
R_{hmSN,crit} = \left(\frac{2}{N^{1/3}}\right)^{\frac{12\gamma}{(15\gamma-11)}}
   \left(\frac{25 G}{7 \beta}\right)^{\frac{9(\gamma-1)}{(15\gamma-11)}}
   \times 
      \nonumber      
          \\[0.2cm]      
          & & \hspace{0.9cm} 
   \left[\frac{9.1 (9\gamma-4)(1 - \epsilon) N}
   {125(\gamma-1) L_{\star}}\right]^{\frac{6(\gamma-1)}{(15\gamma-11)}} 
   \times
      \nonumber      
          \\[0.2cm]      
          & & \hspace{0.9cm} 
\left[\frac{(1-\epsilon) G M_{tot}}
   {\beta (\gamma-1) E_0}\right]^{\frac{4}{(15\gamma - 11)}}
       M_{tot} ,
       \label{eq13a}
      \\[0.2cm]     \label{eq13b}
      & & \hspace{-1.1cm}
\rho_{SN,crit} = \frac{3 M_g}{4 \pi a^3_{crit}} = 
              \frac{3 (1 - \epsilon) M_{tot}}
              {4 \pi (R_{hmSN,crit}/1.3)^3} .
\end{eqnarray}

SN-driven bubbles sweep up the gas left over from star formation and 
if they stall at the distance larger than the mean separation between nearby 
massive stars, the stellar winds formerly separated by the dense matter left 
over from star formation merge and add their energy to the thermal energy of 
the hot bubble formed after a supernova explosion.  A coherent shell 
that sweeps up the gas left over from star formation is then formed inside
the cluster. The energy input rate in such a bubble grows rapidly as it 
overtakes neighboring wind sources. One can show (see Appendix A) that 
in this case the expanding shell accelerates even if it moves into a constant 
density ambient medium. Such an accelerating shell becomes destroyed via RT 
instabilities. 
The hot gas then finally escapes from the cloud carrying away all supernova 
products whereas fragments from the broken shell are likely to remain bound 
within a gravitational well of the cluster \citep[][]{Krause2012}.
Massive stars not affected by the SN blast wave then continue to contaminate 
the collection of RT clumps formed after blowout of previous SNRs as well as 
the rest of the gas left over after the formation of a first stellar 
generation. 
  
In the central zones of even more compact and dense star forming clouds SNRs 
stall before merging with nearby wind-driven bubbles.  Only in this case 
and with the help of the global turbulent pressure SNe are able to eventually
enrich the pristine gas left over from the 1G formation with iron group 
elements. The radius of this zone, $R_{enrich}$, is determined by the critical
half-mass radius $R_{hmSN,crit}$ and the actual half-mass radius of the 
proto-stellar cloud.

\section{Gas expulsion vs gas retention in young stellar clusters}

As mentioned above, some models of multiple populations in globular clusters 
require the star cluster volume to be cleaned up whereas other require the 
primordial gas to be retained. In this section the critical lines which allow 
one to distinguish between clusters which expel or retain the gas left over 
from star formation are presented. Also, how these critical lines are affected
by the proto-stellar gas metallicity and by the SFE are here discussed.

\subsection{Young clusters in present day galaxies}

Figure 1 presents the critical radii, $R_{hmw,crit}$ and $R_{hmSN,crit}$
(upper panels), and the critical gas central densities (lower panels)
as a function of the star forming cloud mass, all assuming a solar gas 
metallicity. The left-hand and right-hand panels in Figure 1
correspond to different star formation efficiencies: $\epsilon = 0.3$ and
$\epsilon = 0.7$, respectively. The critical radii and critical gas central
densities for the wind-dominated and SN-dominated regimes are displayed by 
solid and dashed lines, respectively. 

There are three zones in these diagrams. In clusters located above the solid 
line on the top panels and below the solid lines on the bottom panels, 
individual wind-driven bubbles merge with neighboring sources and form a
collection of dense clumps in the whole star forming volume. Nearby stellar
winds then begin to collide heating up the injected matter and forming a mass
loaded star cluster wind which eventually cleans up the star forming cloud. 
Stellar winds are not able to merge in more compact and denser star forming
clouds located in between the solid and dashed lines in these diagrams. 
However SNRs formed in such clusters engulf neighboring wind sources. They 
gain then more energy which results in their acceleration and eventually leads 
to the development of RT instabilities and the SN shell destruction. 
It is likely that RT clumps left over from the disrupted shells remain bound 
to the cluster \citep{Krause2012} while the iron enriched gas escapes into 
the ambient ISM through the broken shells and does not pollute the gas 
available for a 2G. The fundamental difference between these two regimes
is that individual SN explosions are not synchronized in time and space. 
They do not act coherently and have little effect on the leftover gas 
distribution \citep[][]{TenorioTagle2015}. On the other hand, if wind-driven 
bubbles merge, they do merge simultaneously in the whole star forming volume 
and eventually form a powerful, mass loaded star cluster wind. The coherence 
of energy sources is a fundamental ingredient as was already stressed by 
\citet{Calura2015, Sharma2014}.

The most massive and compact star clusters located below the dashed lines in 
the top panels and above the dashed lines in the bottom panels, retain the 
pristine gas and some supernovae products. In this case the 2G should present
an enhanced iron abundance.

The comparison of the left-hand and right-hand panels in Figure 1 shows that 
the regime of star formation strongly depends on the SFE of the 1G. The 
critical lines go down in the top panels and up in the bottom panels as the 
considered SFE for the 1G becomes larger. This implies that natal clouds with 
a large SFE must be more compact and denser compared to their 
counterparts with a low SFE in order to form the 2G subpopulation. Note also 
that the critical lines mark clouds which retain primordial gas only in 
a small central zone. However, the size of this zone and the amount of the 
retained gas available for the 2G formation grow rapidly in more compact 
clusters with an $R_{hm} < R_{hm,crit}$. 

The critical gas central densities are presented on the bottom panels. At 
first glance they look too large if one takes as a reference value the 
interstellar gas density in the Milky Way or another nearby galaxy. However, 
the stellar mass density in globular clusters with multiple stellar 
populations reaches $10^5$\Msol \, pc$^{-3}$ which corresponds to a number 
density of atoms about $n \approx 10^7$cm$^{-3}$ \citep[][]{Renzini2013}.
This value is comparable and in many cases even larger than the critical 
densities shown in Figure 1. Thus, it is likely that many young, massive and 
compact clusters may retain the gas left over from star formation in their 
central zones. This seems to be in conflict with the recent results by 
\citet{Bastian2013,Bastian2014}
who did not detect a significant amount of gas in many young (ages less than 
20 Myr) massive (masses about $10^6$\Msol \, or more) clusters located in 
different nearby galaxies and claimed that even high-mass clusters expel 
their natal gas within a few Myr after formation. 
Note, however, that not all nearby young massive clusters are free of gas. 
Probably the best example of an equally young and massive star cluster that
still remains embedded within its natal cloud is a $\sim 10^6$\Msol, 
$\sim 4$Myr
old cluster, in the dwarf galaxy NGC 5253  \citep[see][]{Turner2000,Turner2003,
Turner2015,Beck2015}. In order to understand what determines such a profound 
difference between this cluster and those studied by Bastian and 
collaborators, we selected from the list of \citet{Bastian2014} two clusters 
whose ages and masses are similar to the massive cluster in NGC 5253. 
Following \citet{Krause2016} we transformed the half-light radii of the 
selected clusters into half-mass radii (see Table 1). Note that the 
half-mass radii may be even larger if one accounts for a possible mass 
segregation in the observed clusters. As the cluster in NGC 5253 is still 
embedded into a dense molecular cloud and is not visible at optical or UV 
wavelengths, we adopted the size of the radio supernebula detected around the 
cluster \citep[see][]{Turner2000,Beck2015} as the maximum value for the star 
cluster half-mass radius and then accommodated all clusters in our diagram.  
\begin{table*}[htp]
\caption{\label{tab:1} Observational parameters of the selected clusters}
\begin{tabular}{c c c c c}
\hline\hline
Galaxy & Cluster & Age & Mass & Half-mass Radius \\
       &         & (Myr) & ($10^6$\Msol) & (pc)  \\
\hline
\hline 
The Antennae   &                     &      &       &        \\
               &  T352/W38220        & 4    & 0.92  & $4.1^{+2.5}_{-1.7}$    \\
               &  Knot S             & 5    & 1.6   & $13.6^{+2.5}_{-2.5}$   \\
NGC 5253       &  Super star cluster &      &       &        \\
               &  in cloud D         & 4.4  & 1.1   & $0.5^{+0.15}_{-0.15}$ \\
\hline\hline
\end{tabular}
\end{table*}

In the case of a low star formation efficiency (see the left-hand upper
panel in Figure 1) T353/W38220 and NGC 5253 clusters are located in the 
parameter space were clusters should retain gas left over from star 
formation even in the supernovae-dominated regime. Even the older and less
concentrated cluster Knot S is located in between the wind and the SN critical
lines and thus should expel SN products, but retain the primordial gas and
the dense RT clumps caused by blowout events. However in the case of a larger 
star formation efficiency (the right-hand upper panel in Figure 1), only NGC 
5253 cluster remains in the total gas retention parameter space, while the 
Antennae clusters are to disperse their primordial gas during the 
wind-dominated stage. Therefore we suggest that the star formation efficiency 
in the Antennae clusters must have been large. The major difference between 
these three clusters, very similar in their mass and age, is then their 
compactness: the most compact cluster in the NGC 5253 galaxy remains embedded 
into its natal cloud whereas the less compact clusters in the Antennae are 
not. The stalling bubble model may also explain the lack of non-thermal radio 
emission from the NGC 5253 cluster \citep[][]{Beck1996,MartinHernandez2005}
because in this case shock waves vanish rapidly, the Fermi acceleration 
mechanism does not work and therefore high energy particles cannot survive 
for a long time after supernova explosions. 
\begin{figure*}
\vspace{16.5cm}
\includegraphics{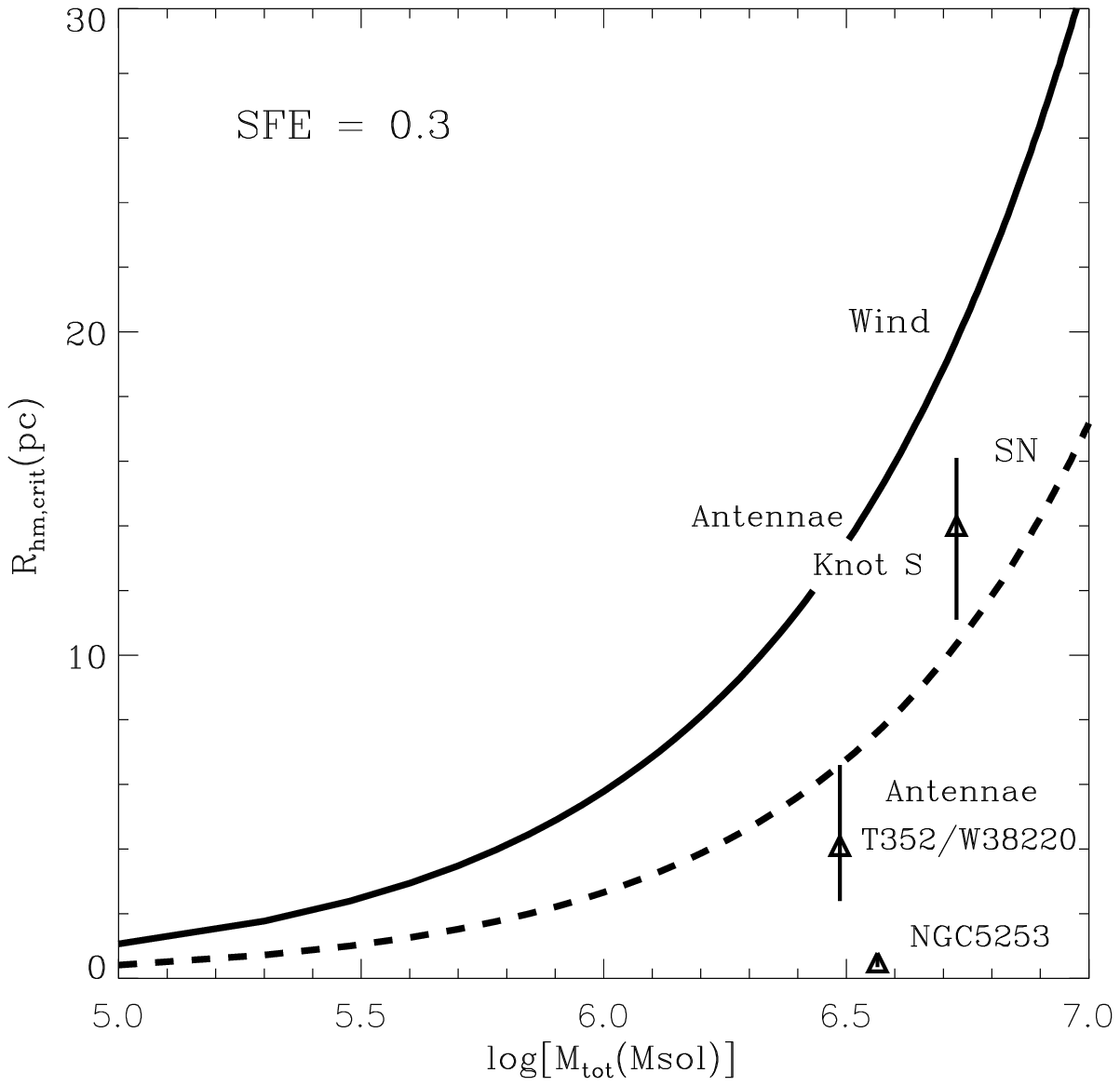}
\includegraphics{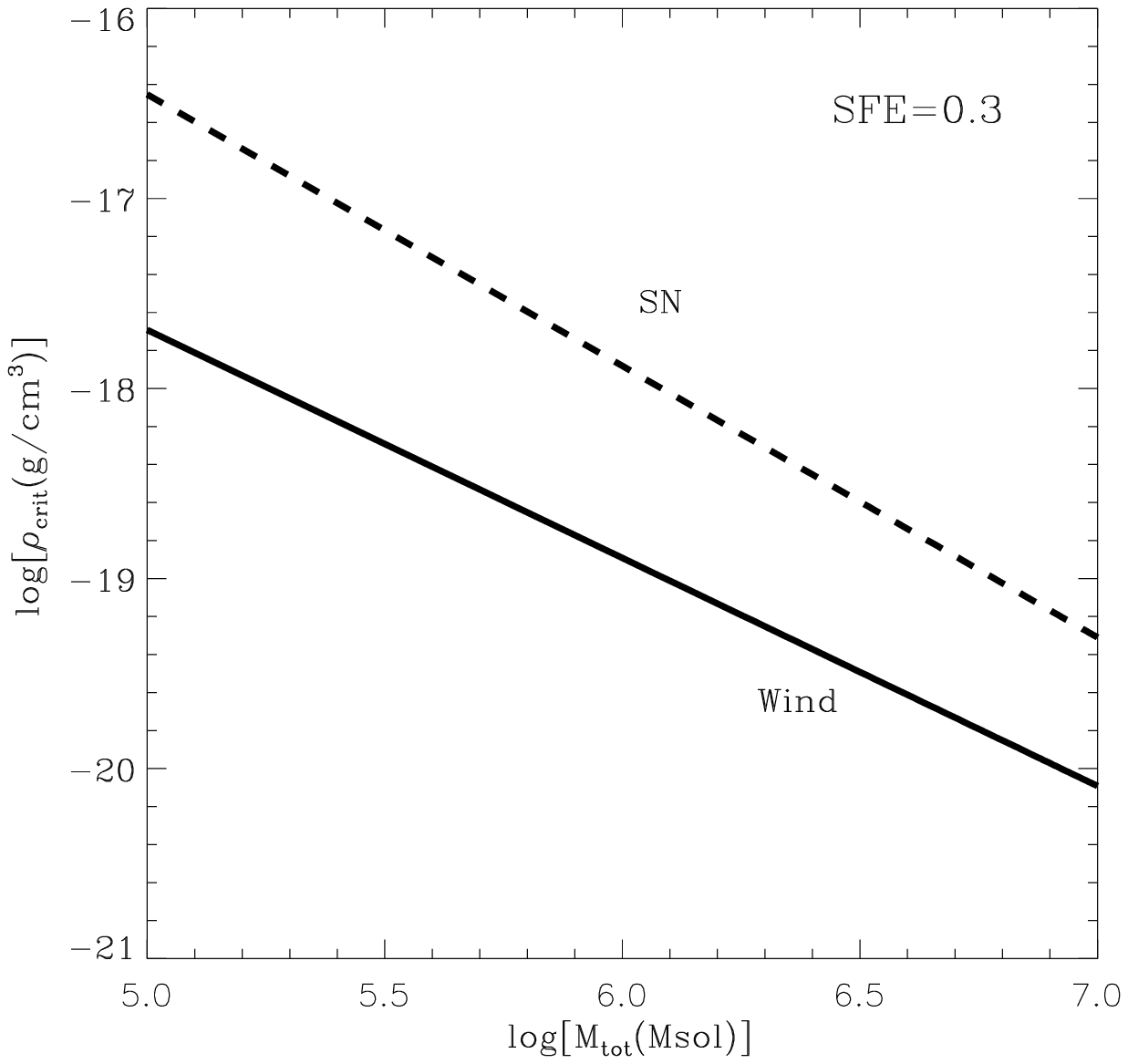}
\includegraphics{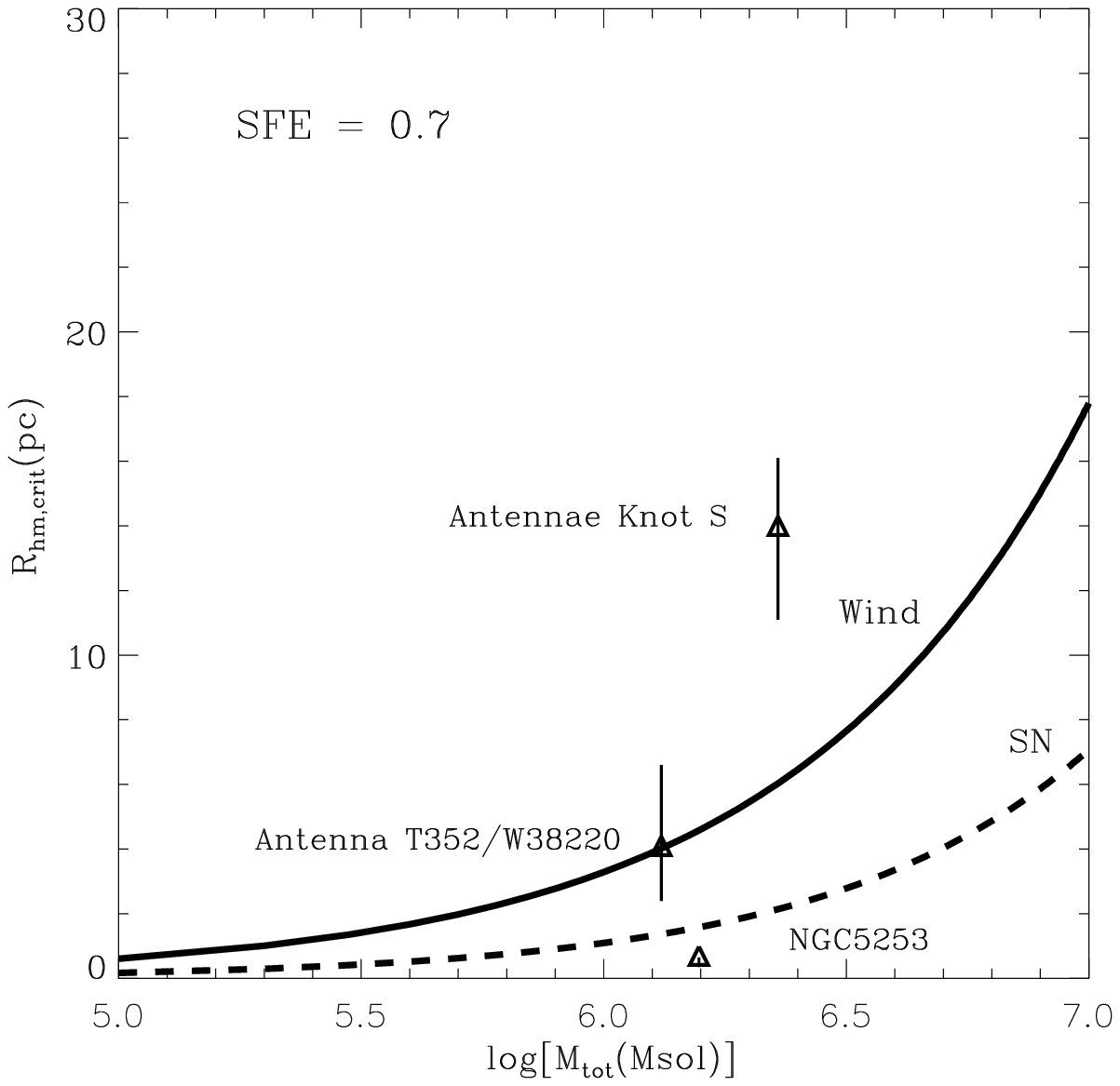}
\includegraphics{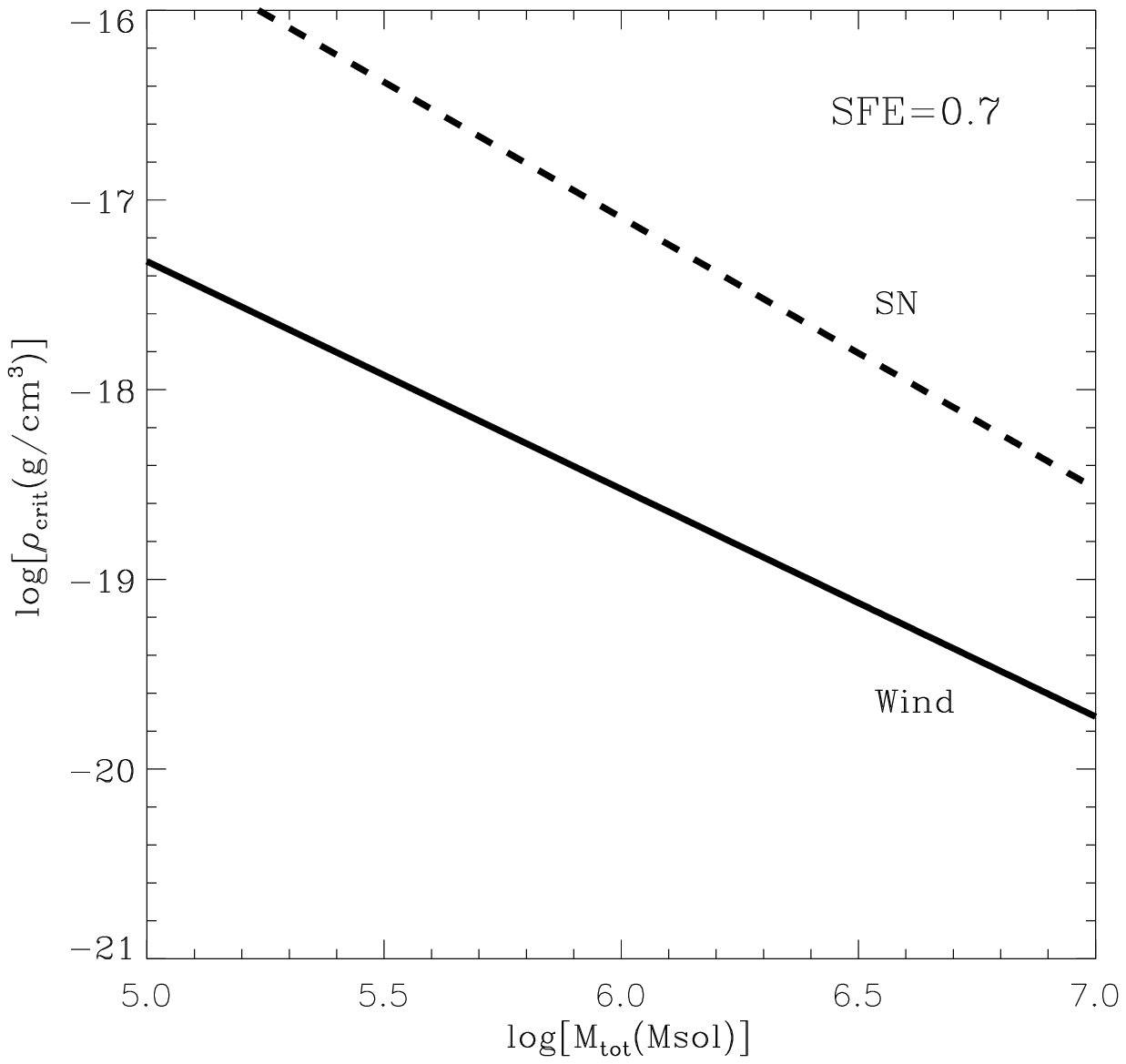}
\caption{Critical radii and gas central densities for proto-stellar 
         clouds with solar gas metallicity. The left-hand panels show how the 
         critical half-mass radius and the central gas density change with 
         the total mass of the proto-stellar cloud if the star formation 
         efficiency $\epsilon = 0.3$. The right-hand column shows the same 
         in the case of the larger star formation efficiency $\epsilon = 0.7$.
         In all panels solid and dashed lines display the critical cloud 
         parameters derived for a stellar wind-dominated regime and
         after the onset of the supernovae explosions, respectively. Labeled
         symbols in the upper panels give the location of three selected 
         clusters (see Table 1) with their corresponding error bars.}
\label{f1}
\end{figure*}

\subsection{Young clusters in ancient low metallicity galaxies}

There is a common belief that stellar winds driven by stars with a low metal 
abundances are less energetic than those driven by their  metal rich 
counterparts. Taking into account this difference, one can calculate 
the critical half-mass radii and the critical central densities for young 
proto-stellar clouds with a low gas metallicity as was done in the 
previous section for present day young massive clusters. Critical lines 
calculated with a stellar wind mechanical luminosities as derived by 
\citet{DErcole2008} for stars with extremely low metal abundances, 
$L_{\star} = 3 \times 10^{39} (M_{\star}/10^6$\Msol) ers s$^{-1}$, 
are shown in Figure 2.
\begin{figure*}
\vspace{16.5cm}
\includegraphics{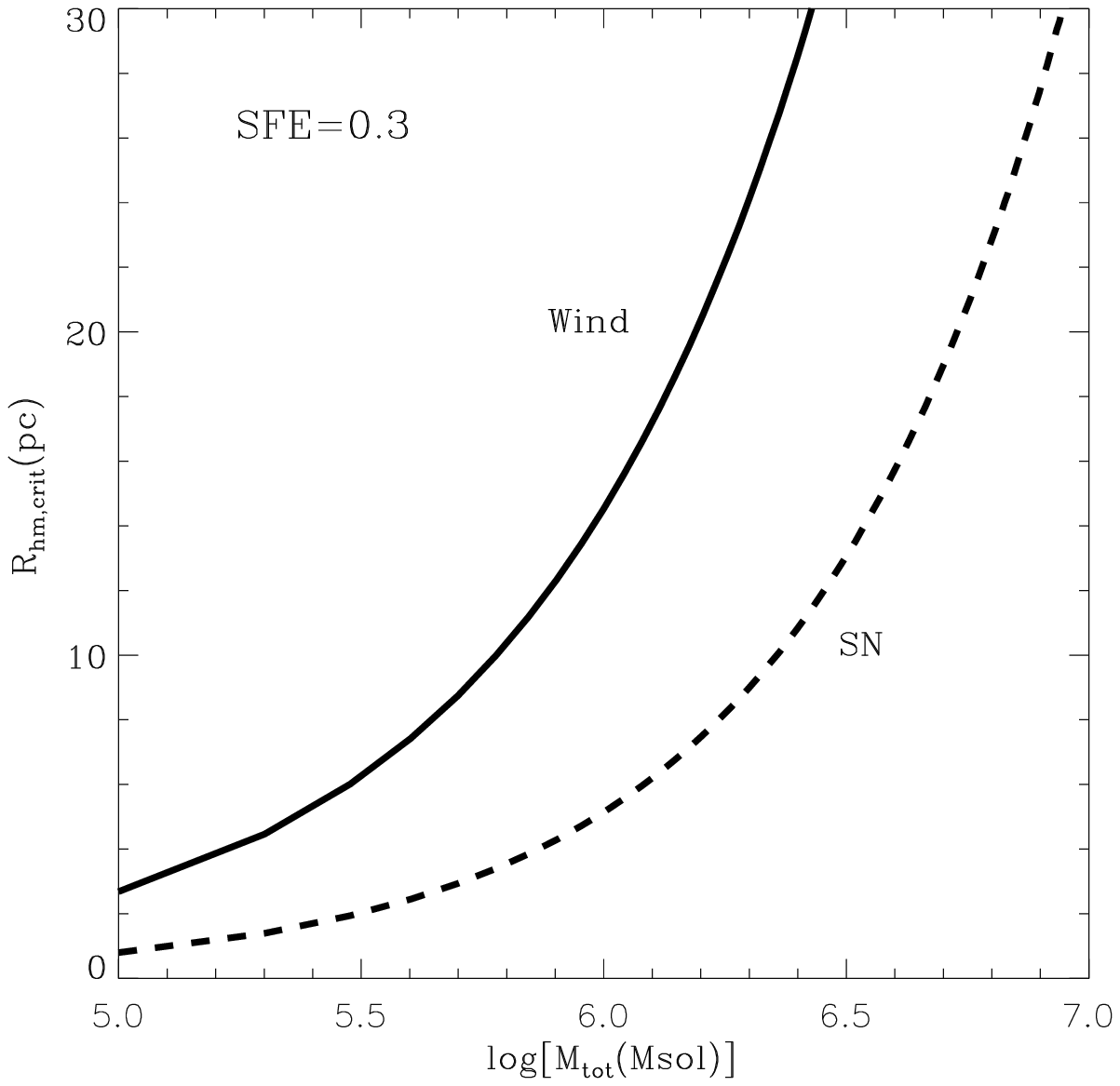}
\includegraphics{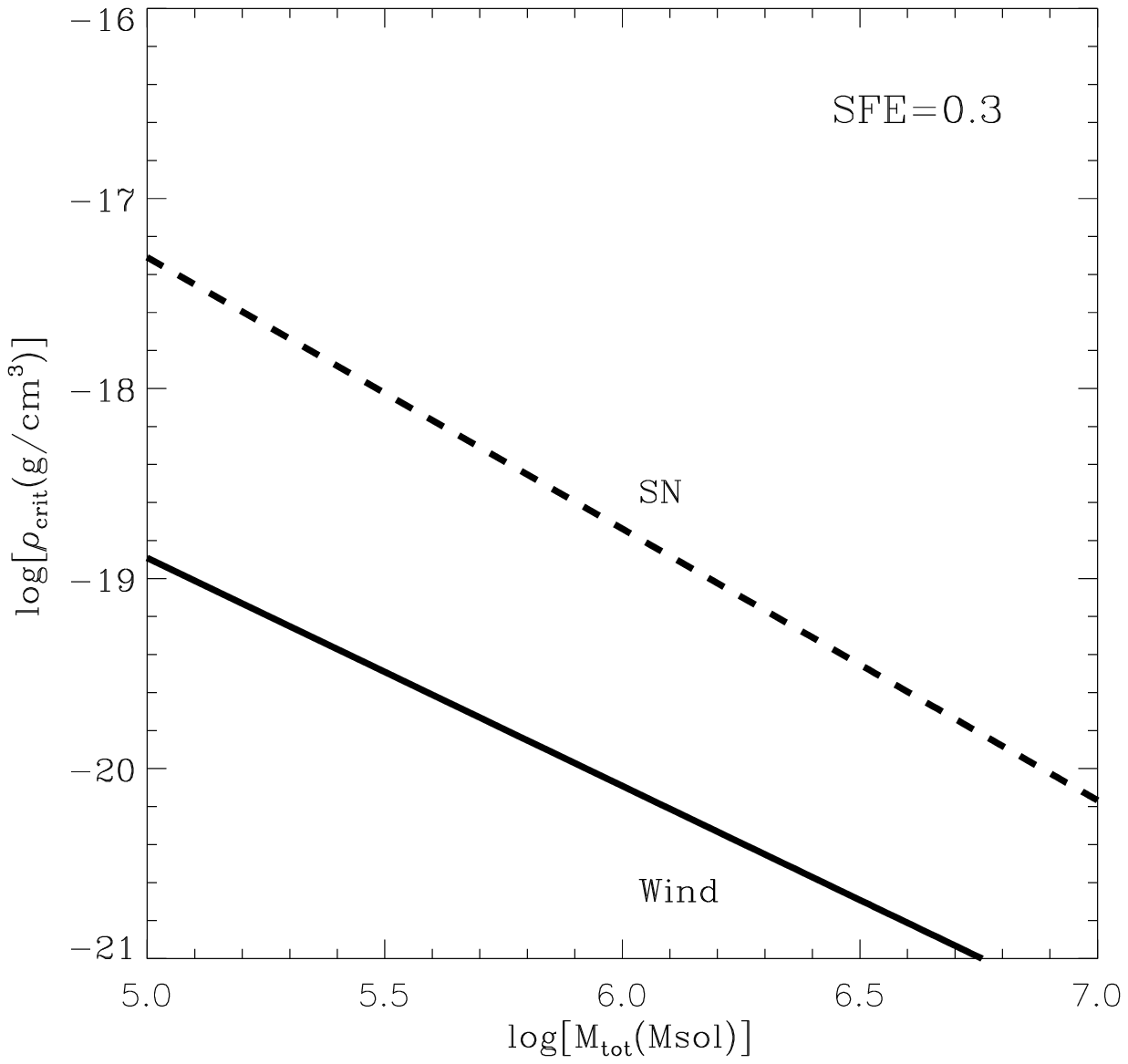}
\includegraphics{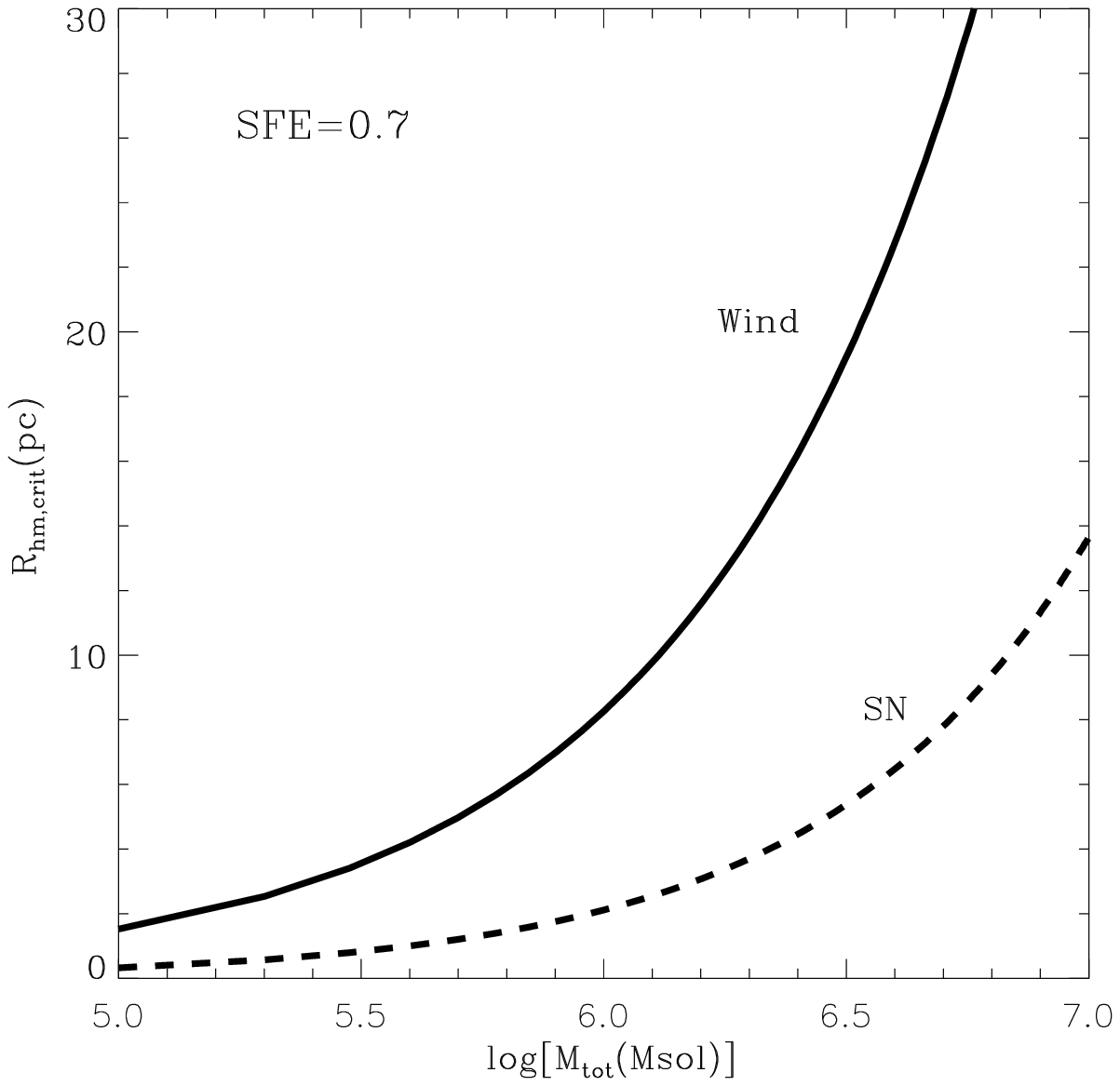}
\includegraphics{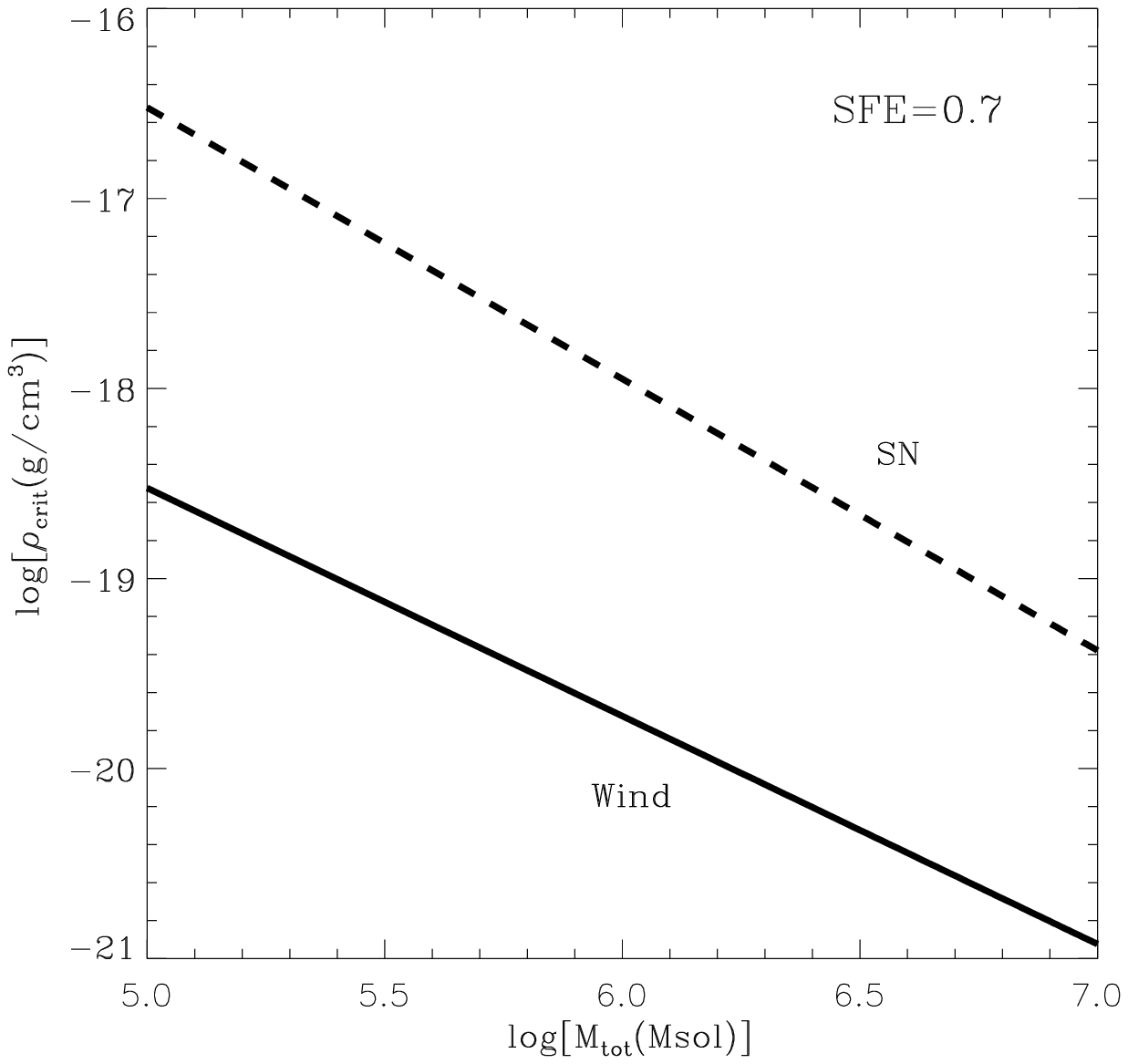}
\caption{The same as in Figure 1, but calculated for proto-stellar clouds 
         with a primordial gas metallicity $Z = 0.001 Z_{\odot}$.} 
\label{f2}
\end{figure*}

An inspection of Figure 2 and its comparison with Figure 1 lead one to 
conclude 
that the requirements for proto-globular clusters to retain pristine gas left 
over from star formation are less stringent than those derived for present
day massive clusters. Much less dense (about an order of magnitude) and less 
compact proto-globular clusters may retain gas left over after the 1G 
stars have formed. This is because stars with a low metal abundances have less 
energetic winds and their wind-driven bubbles stall without merging with 
their neighbors in lower density star forming clouds (compare the bottom 
panels in Figures 2 and 1). In this case SNRs also stall before merging
with nearby stellar winds in less dense star forming clouds as they explode
inside wind-driven bubbles with a smaller size. This significantly extends the
window of opportunity to form a second stellar population in a low-metallicity 
clouds.

Clouds located below the solid lines on the top panels and above the solid 
lines on the bottom panels in Figures 1 and 2 are able to form multiple
stellar populations. The mass ratio $M_{1G}/M_{G2}$ depends on the star 
formation efficiency \citep[see][]{TenorioTagle2016} and on the cluster 
compactness. Proto-globular clouds form more massive 2G population when
located further away from the wind-driven bubble critical line in the gas 
retention parameter space. However only very massive and compact clusters 
located below the dashed lines on the top panels and above the dashed lines 
in the bottom panes in Figures 1 and 2 are able to retain SN products and form
stellar populations with different metallicities regarding the iron group 
elements.

\section{Concluding remarks}

The common belief that stellar winds and SNRs in compact, young stellar 
clusters rapidly merge due to a small separation between nearby massive 
stars and expel the gas left over from star formation into the ambient
ISM should be taken with care. 
Four major parameters (total mass, size, star formation efficiency  and the 
natal gas metallicity) determine which one of the two processes - gas 
expulsion or gas retention - dominates in each star forming cloud and  
select one of the three possible star formation regimes:

- if the central gas density in the star forming cloud is small ($\rho_g < 
\rho_{w,crit}$), the collection of individual wind-driven bubbles 
merge and fragment. Neighboring stellar winds then collide, heat up the
injected matter and form a powerful mass-loaded star cluster wind which
eventually cleans out the cluster. In this case 2G stars are not formed unless 
the cluster accretes a sufficient amount of gas during the late stages of 
its evolution, as suggested in the AGB scenario 
\citep[][]{DErcole2008,DErcole2010} for globular cluster formation;

- if the central gas density falls into the range $\rho_{w,crit} < \rho_g <
\rho_{SN,crit}$, stellar winds do not merge. However shells formed after 
individual supernova explosions engulf neighboring massive stars, gain their 
stellar wind energy, accelerate and eventually are disrupted via RT 
instabilities. As individual SN explosions are not synchronized in time and 
space, they have little effect on the gas distribution in the rest of the 
star forming cloud. RT clumps are likely to remain bound within the 
gravitational well of the cluster. This collection of clumps and the gas 
unused for the 1G formation are continuously contaminated by the 1G stars and 
likely form finally a second subpopulation. The hot, iron enriched gas however
escapes from the cluster into the ambient ISM thorough the broken shells and 
does not pollute the 2G with products of SN explosions. All stellar 
subpopulations in such clusters should present the same iron group 
metallicity;

- if the central gas density in the proto-stellar cloud is large enough,
$\rho_g > \rho_{SN,crit}$, SN remnants in the central zones of the star 
forming cloud stall before merging with nearby stellar winds. Such clusters 
retain metals injected into the intra-cluster medium by SNe. Only in this 
case clusters with an [Fe/H] spread could be formed.

Clusters with the same mass, SFE and the natal gas metallicity but different 
mass concentrations evolve in different hydrodynamic regimes. More compact 
clusters retain the natal gas left over from the 1G formation whereas less 
compact ones expel it into the ambient interstellar medium. This justifies
the presence of natal gas in the most compact super star cluster in NGC 5253 
and the non-detection of gas left over from star formation in the less compact
clusters T352/W38220 and Knot S, in the Antennae galaxies.

\section*{Acknowledgments}

The authors thank J. Turner for her detailed comments regarding the size of 
the NGC 5253 super star cluster and the surrounding molecular cloud 
properties. SS thanks E. Ramirez-Ruiz and colleagues from the Astronomy and 
Astrophysics Department for discussions, support and hospitality during his 
stay at UC Santa Cruz. We also thank our anonymous referee for a detailed
report full of valuable comments which helped to clarify our model and improve 
the paper significantly. This study has been supported by CONACYT - M\'exico, 
research grant 167169.

\bibliographystyle{mnras}
\bibliography{GC}

\appendix

\section{SNRs in star forming clouds pervaded with stalled winds}

The hot bubble formed after a SN explosion should overlap nearby stellar winds 
which did not merge in the dense gas left over from 1G formation. In this case
the wind energies are added to the thermal energy of the bubble. In the 
central zones of the cloud where the gas density is almost homogeneous, the 
equations describing the expansion of such a bubble and its leading shell are: 
\begin{eqnarray}
      & & \hspace{-1.1cm} 
M_{sh} = 4 \pi \rho_g R^3 / 3 ,
       \label{A1}
          \\[0.2cm]      
          & & \hspace{-1.1cm} 
\der{(u M_{sh})}{t} = 4 \pi P R^2 ,
       \label{A2}
          \\[0.2cm]      
          & & \hspace{-1.1cm}
\der{E_{th}}{t} = 4\pi n_{\star} L_s R^3/3 - 4\pi P u R^2 ,
        \label{A3}
          \\[0.2cm]      
          & & \hspace{-1.1cm}
P = (\gamma-1) \frac{3E_{th}}{4\pi R^3} ,
       \label{A4}
          \\[0.2cm]      
          & & \hspace{-1.1cm}
u = \der{R}{t} ,
       \label{A5}
\end{eqnarray}
where $n_{\star}$ is the number density of the 1G massive stars, $L_s$ is 
the mechanical power of a single stellar wind, $R$, $M_{sh}$ and $u$ are 
the radius, mass and the expansion velocity of the shell, respectively. 
$P$, $E_{th}$ are the gas thermal pressure and energy inside the remnant,
and $\rho_g$ is the density of the gas left over from star formation in
the central zones of the cloud. Combining equations (\ref{A2}) and (\ref{A4}) 
one can obtain:
\begin{equation}
      \label{A5}
E_{th} = \frac{R}{3(\gamma-1)} \left[u \der{M_{sh}}{t} + 
                  M_{sh} \der{u}{t} \right] .
\end{equation}
Substituting this equation into (\ref{A3}), excluding $P$ by means of
(\ref{A2}) and taking into account that in the case of a homogeneous
gas distribution ${\rm d}M_{sh}/{\rm d}t = 4 \pi \rho_g u R^2$, one can 
obtain:
\begin{equation}
      \label{A6}
u \der{u}{t} + \frac{3(3\gamma-2)}{3\gamma+4} \frac{u^3}{R} =
               \frac{3(\gamma-1)}{3\gamma+4} \frac{n_{\star} L_s}{\rho_g} .
\end{equation}
This equation has a power-law solution:
\begin{eqnarray}
      \label{A7}
      & & \hspace{-1.1cm} 
R = \left[\frac{8(\gamma-1)n_{\star}L_s}{3(30\gamma-14)\rho_g}\right]^{1/2} 
     t^{3/2}
       \label{A8}
          \\[0.2cm]      
          & & \hspace{-1.1cm} 
u = \frac{3}{2}
    \left[\frac{8(\gamma-1)n_{\star}L_s}{3(30\gamma-14)\rho_g}\right]^{1/2}
    t^{1/2} .
\end{eqnarray}
Thus if the shell formed by a single SN overtakes neighboring wind sources,
it would unavoidably accelerate and blowout even if expands into an ambient 
gas with a homogeneous density distribution.

\bsp	
\label{lastpage}
\end{document}